\documentclass[amsmath,trackchanges,twocolumn]{aastex702}
\shorttitle{Candidate TEO in CZ\,Tucanae}
\shortauthors{Kołaczek-Szymański et al.}
\received{October 2, 2026}
\revised{November 23, 2026}
\accepted{December 1, 2026}
\submitjournal{AAS Journals}

\begin{document}


\title{Eppur Binaria Non \`e Esclusa: Pulsations in the Long Secondary Period Red Giant CZ\,Tucanae Appear to Be Tidally Excited by Its Veiled Companion}

\author[0000-0003-2244-1512]{Piotr A. Kołaczek-Szymański}
\affiliation{Astronomical Observatory, University of Warsaw, Al. Ujazdowskie 4, 00-478 Warsaw, Poland}
\affiliation{Astronomical Institute, University of Wrocław, Mikołaja Kopernika 11, 51-622 Wrocław, Poland}
\email[show]{piotr.kolaczek-szymanski@uwr.edu.pl}
\correspondingauthor{Piotr A. Kołaczek-Szymański}

\author[0000-0002-6212-7221]{Patryk Iwanek}
\affiliation{Astronomical Observatory, University of Warsaw, Al. Ujazdowskie 4, 00-478 Warsaw, Poland}
\email{email@email.pl}

\author[0000-0002-4544-0750]{Jim Fuller}
\affiliation{TAPIR, Mailcode 350-17, California Institute of Technology, Pasadena, CA 91125, USA}
\email{email@email.pl}

\author[0000-0001-9439-604X]{Dorota M. Skowron}
\affiliation{Astronomical Observatory, University of Warsaw, Al. Ujazdowskie 4, 00-478 Warsaw, Poland}
\email{email@email.pl}

\author[0000-0002-6495-0676]{Grzegorz Pojmański}
\affiliation{Astronomical Observatory, University of Warsaw, Al. Ujazdowskie 4, 00-478 Warsaw, Poland}
\email{email@email.pl}

\author[0000-0002-7777-0842]{Igor Soszy{\'n}ski}
\affiliation{Astronomical Observatory, University of Warsaw, Al. Ujazdowskie 4, 00-478 Warsaw, Poland}
\email{email@email.pl}

\begin{abstract}
Long secondary periods (LSPs), observed in about one-third of pulsating red giants, remain unexplained despite decades of study, with binarity and non-radial oscillations as the leading competing hypotheses. If LSP variables are indeed interacting, eccentric binary systems consisting of a red giant and a substellar object, tidally excited oscillations (TEOs) would be expected to appear in a fraction of them. Therefore, detection of TEOs offers an independent test of the binary scenario that has not previously been applied to LSP stars. Using a combined, approximately 80-year light curve assembled from \emph{Digital Access to a Sky Century @ Harvard} (DASCH), Hipparcos, and \emph{All Sky Automated Survey} (ASAS) photometry, we find that CZ Tucanae (CZ\,Tuc) exhibits a pulsation period of $55.4734\pm0.0074$ days along with its LSP of $443.62\pm0.12$ days, with a ratio of $7.9970\pm0.0024$ -- consistent with a TEO. Theoretical models show the observed amplitude of the candidate TEO could be produced after adopting the estimated fundamental parameters of CZ\,Tuc. Our findings identify CZ\,Tuc as the first candidate LSP star with a TEO, providing independent evidence for binarity among at least a subset of these enigmatic variables. In addition, period ratios close to integer values, most notably 7, 8, and 9, are clearly overrepresented among OGLE-III LSP stars in the Magellanic Clouds, showing that CZ\,Tuc is not an isolated case.
\end{abstract}



\section{Introduction}\label{sec:introduction}
The long secondary period (LSP) phenomenon, which affects about 30\% of all pulsating red giants \citep[e.g.,][and references therein]{2022pas..conf..154S}, has remained an unsolved puzzle of observational astrophysics for almost a century \citep{oconnell1933}. Several potential solutions to this widespread phenomenon have been proposed to date, including the following scenarios: (i) oscillatory convective modes excited in the outer, strongly non‑adiabatic layers of the giant's envelope \citep{saio2015,takayama2020}; (ii) the turnover of giant convective cells in the deep convective envelope \citep{1971A&A....10..290S,stothers2010}; (iii) episodic, asymmetric dust formation and mass loss \citep{wood2009,2019MNRAS.484.4678M,2021A&A...649A.110P}; and (iv) a low-mass companion dragging along a co-orbiting dusty cloud \citep{wood1999,soszynski2021}. Unfortunately, despite decades of effort, none of these solutions has managed to fully account for the properties of LSPs as a whole.

The hypothesis that binarity is the source of LSP signals appears to be one of the most promising. In this model, the red giant is orbited by a much less massive (likely sub-stellar) companion on a close orbit with moderate eccentricity \citep[see][and references therein]{decin2025}. The gravitational influence of the companion on the immediate surroundings of the red giant leads to the formation of a dusty ``trailing tail'' that can periodically obscure the disk of the evolved component and modulate the system's total brightness. Although this hypothesis has some weaknesses \citep[raised by, e.g.,][]{nicholls2009,goldberg2024,2026PASP..138e4202S,courtneybarrer2026}, the strongest argument currently in its favor is the presence of ``secondary eclipses'' in the mid-infrared light curves of LSP stars \citep[in the WISE W1 and W2 passbands;][]{soszynski2021}. Nevertheless, this argument does not appear to conclude the ongoing debate. Hence, the need to search for other, independent traces of possible binarity among LSP stars.

The search for, and possible detection of, tidally excited oscillations \citep[TEOs; see][and references therein]{kumar1995,fuller2017} offers such an opportunity, one that has not yet been explored in the context of the LSP puzzle. If LSPs are indeed tight, eccentric binary systems, there is a chance that the periodically varying tidal potential of the elusive companion induces oscillations in the red giant. The defining feature of TEOs, as opposed to non-resonant self-excited modes, is that their frequency ($f_{\rm puls}$, in the inertial frame) is equal to an exact integer multiple ($N$) of the orbital frequency ($f_{\rm orb}$). They therefore obey the unique relation $f_{\rm puls}/f_{\rm orb}=N$, which is observed in many ``heartbeat stars'' \citep[e.g.,][]{welsh2011,2016AJ....151...68K,2020ApJ...888...95G,kolaczek2021,wrona2022b}. Tidally excited oscillations are thus an almost direct signature of binarity.

Long secondary period stars cast onto the period\,--\,luminosity diagram of red giants form so-called ``sequence~D'', which overlaps with ``sequence~E'', consisting of ellipsoidal and eclipsing binaries plotted at their orbital periods \citep{wood1999, soszynski2004, derekas2006}. This overlap hints at a common binary origin, yet the two groups differ in radial-velocity amplitudes, eccentricities, and mid-infrared excesses \citep{Nicholls2010}. A TEO in an LSP star would reveal tidal interaction with a companion too low in mass to produce notable ellipsoidal variability \citep{soszynski2007a}, further strengthening the suggestion that sequence~D is a continuation of sequence~E toward lower companion masses.

While inspecting the phase-folded light curves of LSP stars, we noticed by eye that in CZ\,Tuc the pulsation signal phases almost perfectly with the star's LSP. This means that the ratio of the periods of these two signals is very close to an integer, which may indicate the presence of a TEO. In this paper, we present a detailed analysis of the CZ\,Tuc's archival light curve from this perspective. We argue that this is the first LSP with a candidate TEO, which constitutes a further argument in favor of the binarity hypothesis as the origin of the LSP phenomenon.
 
\section{CZ\,Tucanae} \label{sec:cz_tuc}
CZ\,Tucanae (CZ\,Tuc; HD\,6550; Gaia DR3 4690894806035949184) is a Johnson $V\sim9$\,mag star, located in the angular vicinity of the Small Magellanic Cloud and the NGC\,362 globular cluster. However, both its apparent brightness and its geometric distance of approximately 1.26\,kpc \citep{2021AJ....161..147B} confirm that it is a Galactic foreground object. \cite{1975mcts.book.....H} classified CZ\,Tuc as an M2/3\,III spectral-type red giant star. Space-borne observations collected by the Hipparcos satellite in the early 1990s established that the dominant photometric variability of CZ\,Tuc occurs with a period of around 411\,d and a peak-to-peak amplitude of approximately 0.6\,mag in the $H_{\rm p}$ passband \citep{1997A&A...323L..49P}. At the same time, the star was tagged as a semi-regular variable. About a decade later, \cite{2002AcA....52..397P} noticed an additional periodic variation in CZ\,Tuc's light curve with a much shorter period of $\sim$\,47.2\,d -- most likely caused by pulsations. Recently, \cite{2026PASP..138e4202S} suggested that CZ\,Tuc may be, in fact, an LSP variable based on its overall photometric properties and the corresponding \textit{Gaia} DR3 light curve. \cite{2026ApJ..1008..154I} confirmed the LSP status of CZ\,Tuc by inspecting its All Sky Automated Survey (ASAS) data, which are of much better photometric and temporal-sampling quality than \textit{Gaia} DR3 counterpart. The authors reported the LSP equal to $P_{\rm LSP}\approx 443.5$\,d.

Fortunately, CZ\,Tuc is also included in the \textit{Gaia} Focused Product Release (FPR) catalog of Long Period Variables \citep[LPVs;][]{2023A&A...680A..36G} and therefore has a radial velocity (RV) curve. In Appendix~\ref{app:rv}, we show that the dominant variability in the RV curve of CZ\,Tuc is consistent with its LSP, while the amplitude and moderately non-sinusoidal shape fit the general characteristics of LSPs in this regard \citep[e.g.,][]{hinkle2002,nicholls2009}. Fitting a Keplerian orbit to the RV curve indicates an eccentric orbit with $e\approx 0.3$ (Appendix~\ref{app:rv}). Although CZ\,Tuc is listed in numerous photometric and photometric-variability catalogs, apart from \cite{2026ApJ..1008..154I}, its long-term variability on decadal timescales has not yet been studied in detail.

\section{Century-long light curve of CZ Tuc}\label{sec:light_curve}
As we argue in Sect.~\ref{sec:introduction}, the emblematic feature that distinguishes a TEO from an ``ordinary'', non-resonant oscillation mode is its rest-frame (i.e., observed) frequency being exactly an integer multiple of the orbital frequency. The success of establishing this fact ultimately rests on our ability to measure this ratio and, consequently, $f_{\rm puls}$ and $f_{\rm orb}$ with the highest precision possible. In the case of LSP stars, the former frequency typically corresponds to a period of several dozen days, making its measurement relatively straightforward. In contrast, some serious problems can arise with the execution of the analogous task for $f_{\rm orb}$. If the LSP in CZ\,Tuc is also the orbital period (that is, $f_{\rm orb}\equiv f_{\rm LSP}$), then it exceeds one year. Achieving the required sub-percent level of uncertainty in measuring $f_{\rm LSP}$ means that the light curve must span at least several dozen LSP cycles, which translates into several decades. For this reason, we focus on building a combined light curve of CZ\,Tuc with the longest possible baseline.

\begin{figure*}[ht!]
\includegraphics[scale=1]{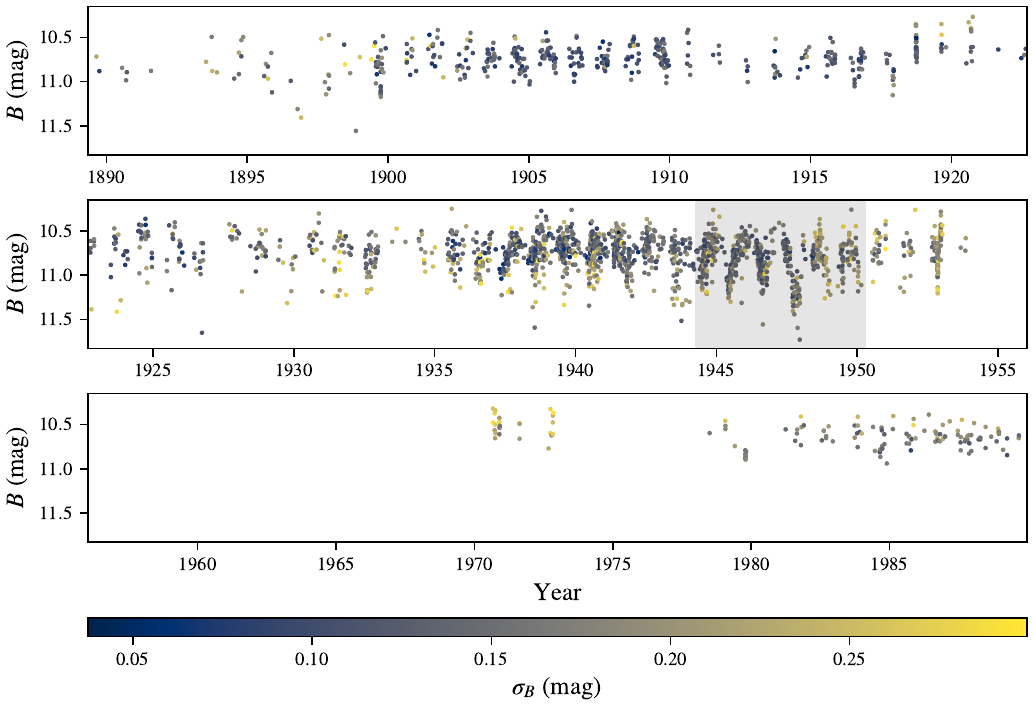}
\caption{Full light curve of CZ\,Tuc from DASCH DR7 after removing faulty and uncertain measurements. The color scale corresponds to photometric errors. To aid visibility, we divided the light curve into three parts of equal time span, shown in the three panels with the same $B$ magnitude range. The gray shaded area in the middle panel indicates the fragment of the light curve where a clear LSP is visible and which we included in the combined light curve (cf. Fig.~\ref{fig:combined_lc}). The LSP signal does not seem to be pronounced in the remaining part of the DASCH time series.}
\label{fig:dasch_full_lc}
\end{figure*}

\begin{figure*}[ht!]
\includegraphics[scale=1]{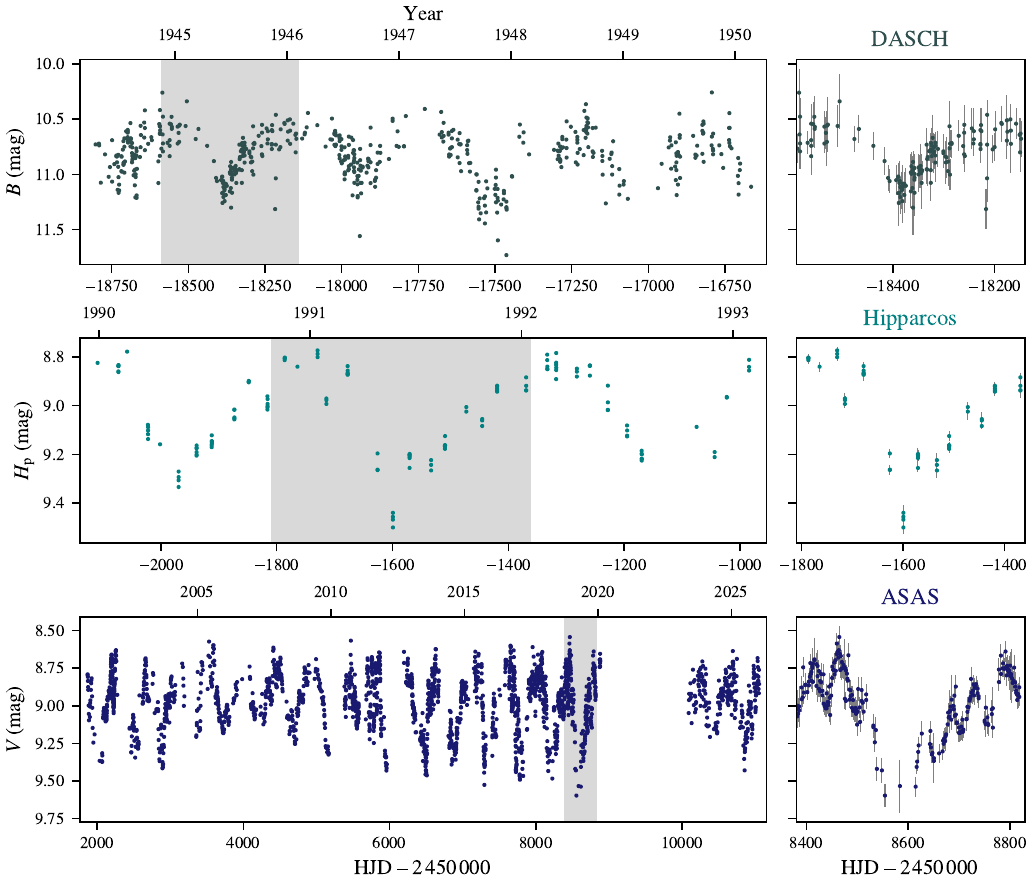}
\caption{Combined light curve of CZ\,Tuc from the DASCH (top row), Hipparcos (middle row), and ASAS (bottom row) surveys. The left column shows the full available light curve. The gray shaded area indicates one selected LSP cycle, which is shown in the right column with the photometric errors. We note that the magnitude scale is the same for both panels in a given row, but differs between rows.}
\label{fig:combined_lc}
\end{figure*}

\subsection{DASCH Project}\label{sec:dasch}
The Digital Access to a Sky Century @ Harvard (DASCH) project\footnote{\url{https://dasch.cfa.harvard.edu/}} was a herculean effort to digitize the world's largest collection of astronomical glass photographic plates, assembled by the Harvard College Observatory \citep{2012IAUS..285...29G}. The greatest advantage of these archival observations is that they span an entire century, from approximately 1880 to 1990. The project was completed after 20 years of work in 2024, culminating in the final DASCH Data Release 7, in which the DASCH Team digitized approximately 428\,000 photographic plates and made more than $3.2\times 10^{10}$ individual magnitude measurements available to the community. The typical photometric error of this dataset is on the order of 0.1\,mag.

Using the interactive \texttt{daschlab} tool \citep{williams_2024_14574817}, we retrieved the DASCH DR7 light curve of CZ\,Tuc in the Johnson $B$ passband \citep{2010AJ....140.1062L,2013PASP..125..857T}. Luckily, our target lies near the SMC in the sky, and is therefore covered by 4\,551 plates spanning the years 1889\,--\,1989. Since photographic plates constitute an extremely heterogeneous dataset, it is necessary to rigorously remove any measurements that may be flawed, either due to defects in the plates themselves or to subsequent digitization and reduction. Following the DASCH documentation\footnote{\url{https://dasch.cfa.harvard.edu/dr7/reduce-lightcurve/}}, we applied the following recommended quality cuts. First, we used \texttt{daschlab}'s built-in \texttt{apply\_standard\_rejections} procedure, which automatically rejects the most problematic measurements. We also excluded all plates with the lowest angular resolution (from the so-called ``meteor campaign''), which are prone to source blending and inaccurate astrometry. Next, we removed all measurements for which the astrometric solution for CZ\,Tuc differed by more than 10\arcsec\ from its actual coordinates. Since the amplitude of the LSP in CZ\,Tuc is $\sim$0.6\,mag, we rejected points with a photometric uncertainty greater than 0.3\,mag. Furthermore, to ensure that CZ\,Tuc's photometry was performed on plates with adequate focus and exposure time, we omitted measurements with $\texttt{ellipticity}>0.5$, as well as all plates with a limiting magnitude lower than $B=12$\,mag (CZ\,Tuc has mean $B\sim 10.7$\,mag). The resulting cleaned light curve contains 2\,245 individual measurements and is presented in Fig.~\ref{fig:dasch_full_lc}.

It is difficult to determine unambiguously whether this is an intrinsic property of CZ\,Tuc or an effect related to the varying quality of the DASCH data, but its LSP variability is well-pronounced only over a relatively short interval of years, $\sim$\,1944\,--\,1950. A possible ``disappearance'' of the LSP signal is not unprecedented among variables of this type, as demonstrated, for instance, by \cite{soszynski2021} and \cite{2026ApJ..1008..154I}. However, since our primary goal is to obtain an accurate $f_{\rm LSP}$, we included in the combined light curve of CZ\,Tuc only the aforementioned fragment of the DASCH time series, between HJD 2\,431\,180 and 2\,433\,400, which is shown in the upper row of Fig.~\ref{fig:combined_lc}.

\subsection{Hipparcos}\label{sec:hipp}
The Hipparcos satellite, operated by ESA between 1989 and 1993, was {\it Gaia}'s predecessor -- primarily designed for high-precision astrometry, but it also provided photometric measurements in the broad-band $H_{\rm p}$, $V_{\rm T}$, and $B_{\rm T}$ system. Each target was observed at multiple irregularly spaced epochs, typically yielding $\sim$100 individual magnitude measurements \citep{1997ESASP1200.....E}. The resulting epoch photometry provides sparsely sampled light curves over the Hipparcos mission lifetime, which can be used to identify long-term variability and, when sufficiently well sampled, to constrain periodic signals.

We downloaded the Hipparcos light curve of CZ\,Tuc directly from the CDS/VizieR service\footnote{\url{http://vizier.cds.unistra.fr/viz-bin/VizieR?-source=I/239}}. Initially, the light curve contained 154 points, but we rejected all those with a quality flag other than ``0'', which may indicate potential problems with the photometry. Our final Hipparcos light curve contains 138 measurements and is depicted in the middle row of Fig.~\ref{fig:combined_lc}. In addition to $H_{\rm p}$-band photometry, CZ\,Tuc also has a similar number of magnitudes from the Tycho catalog in the $B_{\rm T}$ and $V_{\rm T}$ filters, but these are of significantly lower quality. Since their scatter dominates over the variability of the star itself, we did not include these two additional datasets into the combined light curve.

\subsection{ASAS}\label{sec:asas}
The All Sky Automated Survey is a long-term photometric monitoring project initiated in 1997 at Las Campanas Observatory in Chile and later extended with a northern station at Las Cumbres Observatory on Haleakala, Hawaii. The survey was designed to monitor the variability of relatively bright stars over large areas of the sky and has provided photometric time series spanning more than two decades. Over the course of the project, the instrumental setup and observing strategy were upgraded several times, improving both the saturation limit and survey depth. The ASAS cameras cover a very wide field of view of approximately $10^\circ \times 10^\circ$, with a spatial scale of about 10\,--\,15 \arcsec/pixel. As a consequence, blending can significantly affect the photometry in crowded stellar fields, while additional systematic uncertainties may arise from vignetting and field-dependent photometric zero points. Despite these limitations, the long time baseline and dense temporal coverage make ASAS data particularly valuable for studies of long-term stellar variability. More details on instruments, data acquisition, and the reduction pipeline can be found, for instance, in \cite{pojmanski1997, pojmanski2001, pojmanski2002}.

In this paper, we used the ASAS light curve for CZ\,Tuc published by \cite{2026ApJ..1008..154I} and supplemented it with additional post-COVID-19 observations. The final light curve is shown in the bottom row of Fig.~\ref{fig:combined_lc} and consists of 2905 epochs spanning over 25 years.

\subsection{Final remarks on the combined light curve}\label{sec:remarks_lc}
For CZ\,Tuc, public light curves from the ASAS-SN project \citep{2017PASP..129j4502K} and {\it Gaia} DR3 \citep{2023A&A...674A...1G} are also available. However, they overlap in time with the ASAS observations and do not offer a better photometric quality than the latter. Moreover, the brightness of CZ\,Tuc is close to the saturation limit in ASAS-SN ($V\sim10$\,--\,$11$\,mag), while the {\it Gaia} light curve contains relatively few data points (about 30 unique measurements in the $G$ band). For these reasons, we did not include these light curves in our further analysis, as doing so would not have led to a significant improvement in the determination of the LSP. CZ\,Tuc was also observed by the Transiting Exoplanet Survey Satellite \citep[TESS;][]{2015JATIS...1a4003R} in ten sectors grouped into four observing seasons. The TESS photometry shows semiregular pulsations with cycle-to-cycle variations in the times of maximum, typical of red giants of this type and consistent with the stochastic nature of most of the pulsational power seen in the ASAS data (Sect.~\ref{sec:teo}). Therefore, including the TESS data would not offer any qualitative change compared to the ASAS observations. Moreover, both the pulsation period and the LSP of CZ\,Tuc ($\sim$55\,d and $\sim$444\,d, respectively; Sect.~\ref{sec:pre-whitening}) are longer than a single TESS sector ($\sim$27\,d). This makes it difficult to properly stitch the data from separate sectors into a reliable light curve, because instrumental offsets and trends between sectors cannot be easily disentangled from the intrinsic long-period variability. In addition, the TESS data consist of short, widely separated segments, which are not suitable for constraining the LSP or for separating the coherent signal at $8f_{\rm LSP}$ from the stochastic variability. For these reasons, we did not include the TESS light curve in our analysis.

The resulting combined and final light curve (Fig.~\ref{fig:combined_lc}) thus spans approximately 80 years, corresponding to $\sim$66 LSP cycles in CZ\,Tuc. We emphasize that it consists of three light curves in three different passbands, with the average brightness of CZ\,Tuc differing between them. In this context, the Hipparcos and ASAS light curves are quite similar, whereas the DASCH light curve contains $B$-band magnitudes and therefore clearly stands out from the other two. To mitigate this issue, before combining the light curves and performing the prewhitening, we subtracted the median magnitude from each light curve in its respective passband. We note that this does not resolve the problem of potential differences in the amplitudes of LSP and pulsation signals depending on the wavelength \citep{2003aahd.conf..421H,takayama2015}. Fortunately, since we are interested solely in the periodicity of the signals in the combined light curve, accounting for this effect was not necessary. All photometric and RV data used in this work are publicly available from the sources cited in Sects.~\ref{sec:dasch}--\ref{sec:asas} and Appendix~\ref{app:rv}.

\section{Fourier analysis of the light curve}\label{sec:pre-whitening}

\begin{figure*}[ht!]
\includegraphics[scale=1]{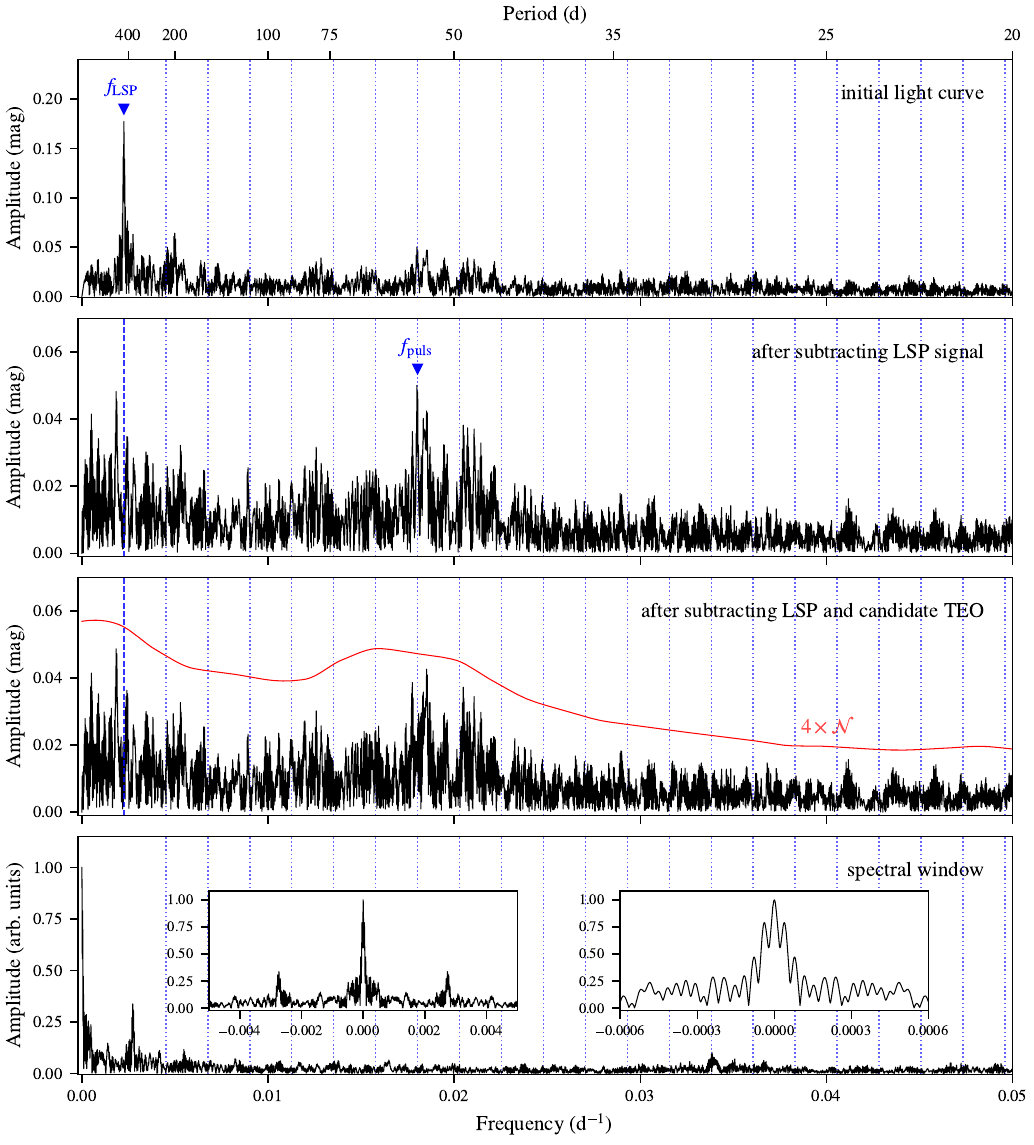}
\caption{Fourier frequency spectra at different stages of prewhitening of the combined light curve, together with its spectral window. (top) Frequency spectrum of the input light curve. The blue triangle marks the frequency corresponding to the LSP. (upper-middle) Frequency spectrum after prewhitening on $f_{\rm LSP}$. The blue triangle indicates the frequency of the dominant pulsation, which is a candidate TEO. (lower-middle) Frequency spectrum after prewhitening on $f_{\rm LSP}$ and $f_{\rm puls}$. The red curve indicates the detection threshold, equal to four times the local noise level. (bottom) Spectral window of the combined light curve, normalized to unity at zero frequency. The inset on the left zooms into the central peak with a pair of annual aliases, and the inset on the right shows the detailed shape of the main peak. In all panels, the harmonics of $f_{\rm LSP}$ are marked by vertical dotted lines. In the upper-middle and lower-middle panels, the vertical dashed line indicates the position of the subtracted $f_{\rm LSP}$.}
\label{fig:prewhitening}
\end{figure*}

To retrieve $f_{\rm LSP}$ and $f_{\rm puls}$, we analyzed the periodic signals in the combined light curve of CZ\,Tuc (hereafter simply referred to as ``the light curve'') using the customary, iterative Fourier-based prewhitening procedure. To calculate the frequency spectra, we used the \texttt{fnpeaks} code\footnote{\url{http://helas.astro.uni.wroc.pl/deliverables.php?lang=en&active=fnpeaks}} (written by W. Hebisch, Z. Kołaczkowski and G. Kopacki), while the periodic variability was modeled with a truncated Fourier series, fitted by the error-weighted least-squares method. At each step of the prewhitening process, we added a sinusoid corresponding to the frequency with the highest signal-to-noise ratio to the variability model. Since the frequency spectra of CZ\,Tuc reveal the presence of red noise due to stochastic variability and (possibly) instrumental effects, we estimated the noise level ($\mathcal{N}$) locally as the mean within a sliding window of 0.01\,d$^{-1}$. We adopted a detection threshold of four times the local noise level, $4\times \mathcal{N}$. We estimated the errors in the resulting frequencies, amplitudes, and phases using the covariance matrix. The spectral window of the light curve is shown in the bottom panel of Fig.~\ref{fig:prewhitening}. Apart from the obvious annual aliases, a weak monthly alias (at $\approx$0.034\,d$^{-1}$), and the daily aliases outside the plotted frequency range, it is free of other significant features.

\subsection{The long secondary period}\label{sec:LSP}

\begin{figure}[ht!]
\includegraphics[scale=1]{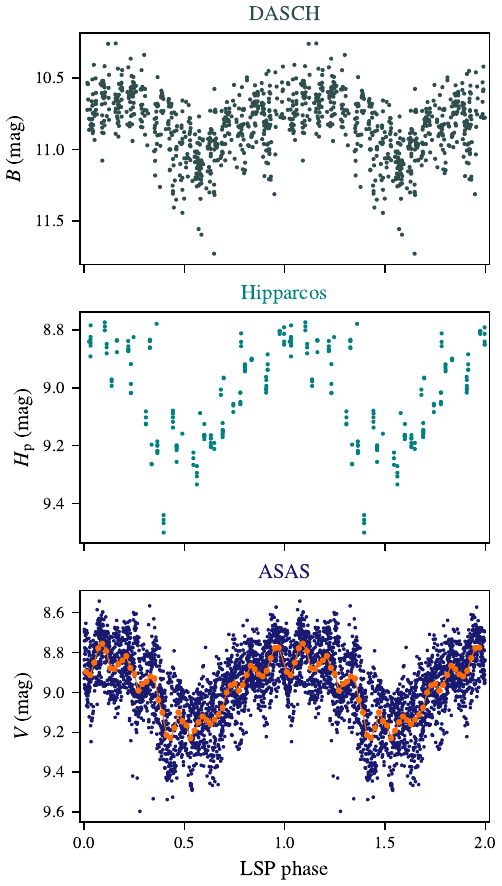}
\caption{Light curves from DASCH (top), Hipparcos (middle), and ASAS (bottom) phased with the LSP of 443.62\,d obtained by us. The zero phase corresponds to HJD\,2\,450\,000. Dark orange dots represent the ASAS light curve binned in phase intervals of 0.02 and reveal the presence of a TEO candidate.}
\label{fig:phased_lcs}
\end{figure}

The frequency spectrum of the input light curve is shown in Fig.~\ref{fig:prewhitening} (top panel). Although we present a frequency range up to 0.05\,d$^{-1}$, we note that no statistically significant periodic signals are present at higher frequencies. Figure~\ref{fig:prewhitening} reveals the clear presence of the frequency corresponding to the LSP, $f_{\rm LSP}$. Although the LSP signal exhibits some degree of non-sinusoidality (cf. Fig.~\ref{fig:combined_lc}), we did not detect any statistically significant low-order harmonics of $f_{\rm LSP}$ in the frequency spectrum. This may be due to elevated noise in this frequency range and/or variations in LSP amplitude and shape. Fitting a sine yields the best-fit $f_{\rm LSP}=0.00225418(61)\,{\rm d}^{-1}$. This corresponds to a period of $P_{\rm LSP}=443.62\pm0.12\,$d. For comparison, the analytical estimate for a coherent sinusoidal signal in white noise, $\sigma_f=\sqrt{6/N_{\rm obs}}\,(\pi T)^{-1}\,\sigma/A$ \citep[][see also \citealt{2025asno.book.....B}]{1999DSSN...13...28M}, where $N_{\rm obs}$ is the number of measurements, $T$ is the total time span of the light curve, $\sigma$ is the rms scatter of the residuals, and $A$ is the semi-amplitude of the signal, yields $\sigma_f\approx3.2\times10^{-7}$\,d$^{-1}$ for the combined light curve ($N_{\rm obs}=3\,600$, $T\approx82$\,yr, $\sigma\approx0.134$\,mag) and the LSP semi-amplitude $A=0.183$\,mag. Our formal uncertainty is thus of the same order and somewhat more conservative. The ASAS, Hipparcos, and DASCH light curves phased with this period are shown in Fig.~\ref{fig:phased_lcs}. There are no obvious changes in the phase of the LSP signal or in its overall morphology over the past 80 years. This supports the assumption of a coherent (phase-stable) LSP signal, which underlies the uncertainty estimates above. Otherwise, the actual uncertainty of $f_{\rm LSP}$ would be larger.

\subsection{The candidate TEO}\label{sec:teo}
\begin{figure}[ht!]
\includegraphics[scale=1]{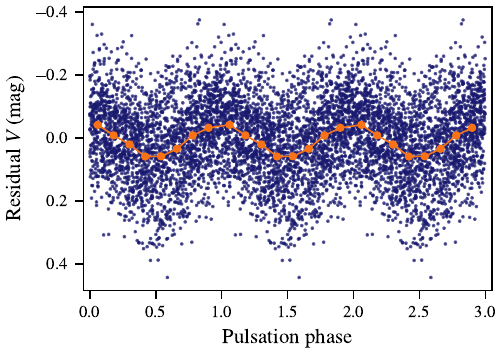}
\caption{ASAS light curve (blue points) after subtracting the LSP signal, phased with $P_{\rm puls}=55.4734$\,d. Phase zero corresponds to the epoch of maximum brightness at HJD\,2\,451\,483.61. The dark orange points with line show the phase-binned light curve, with a bin width of 0.12.}
\label{fig:isolated_teo}
\end{figure}

After prewhitening the light curve on $f_{\rm LSP}$, the residuals reveal the highest maximum at a frequency that lies almost exactly at $8f_{\rm LSP}$ (Fig.~\ref{fig:prewhitening}, upper-middle panel). This signal is also discernible directly in the phased ASAS light curve as an additional photometric wave superimposed on the LSP variability and phase-locked to it (Fig.~\ref{fig:phased_lcs}, bottom panel). Adding this independent frequency to the variability model yields $f_{\rm puls}=0.0180267(24)\,{\rm d}^{-1}$ (translating to the period of $P_{\rm puls}=55.4734\pm0.0074$\,d). We interpret this additional sinusoidal signal as a pulsation in CZ\,Tuc. As in the case of $f_{\rm LSP}$, this formal uncertainty assumes that the signal is coherent, as expected for a TEO. If the signal were (at least partly) due to a stochastically excited mode, the actual uncertainty would be larger (see below). Most importantly, the ratio $f_{\rm puls}/f_{\rm LSP}=7.9970\pm0.0024$ or, equivalently, $|f_{\rm puls}/f_{\rm LSP}-8|=0.0030\pm0.0024$. It is therefore almost exactly equal to 8, within just $1.25\sigma$. This unique feature leads us to conclude that $f_{\rm puls}$ is most likely a TEO with $N=8$. Figure~\ref{fig:isolated_teo} shows the isolated signal of the candidate TEO. The peak-to-peak amplitude of this signal is approximately 0.06\,mag in the $V$ passband. Due to the larger scatter in the measurements, this variability is not detectable in the DASCH and Hipparcos light curves alone.

After subtracting the LSP signal and the TEO candidate from the light curve, the remaining residuals show no evidence of any significant periodic variability (Fig.~\ref{fig:prewhitening}, lower-middle panel). However, apart from the expected increase in noise level toward low frequencies, most of the power around $f_{\rm puls}$ remains in the form of a broad ``bump'' of peaks clustering around a period of $\sim$50 days. Such a power excess is typical of pulsating red giants of this type and is a signature of damped, stochastically excited oscillations \citep[e.g.,][]{2003Ap&SS.284...61B,2005MNRAS.361.1375B,2006MNRAS.372.1721K,Cadmus_2024}. Similar features also emerge in three-dimensional radiation-hydrodynamic simulations \citep[e.g.,][]{2025A&A...699A.148A}. Therefore, the pulsational variability of CZ\,Tuc appears to consist of a ``normal'', semiregular (damped) oscillation and an additional coherent component, which coincides with the highest peak of the bump, located at $8f_{\rm LSP}$. In Sect.~\ref{sec:discussion}, we argue that this is unlikely to be a coincidence and can be explained by a TEO.

\begin{deluxetable}{lccl}
\tablewidth{0pt}
\tablecaption{Summary of CZ\,Tuc's key parameters. \label{tab:summary}}
\tablehead{
\colhead{Parameter} & \colhead{Value} & \colhead{Unit} & \colhead{Reference}
}
\startdata
Mean Johnson $V$ & $9.0$ & mag & this work \\
Geometric distance & $1260^{+27}_{-30}$ & pc & \cite{2021AJ....161..147B} \\
Spectral classification & M2/3\,III & --- & \cite{1975mcts.book.....H} \\
$T_{\rm eff}$ & $\sim3\,550$ & K & \cite{2025ApJ...984...58H} \\
$R$ & $\sim140$ & R$_\odot$ & this work \\
$L$ & $\sim2\,700$ & L$_\odot$ & this work \\
$P_{\rm LSP}$ & $443.62\pm0.12$ & d & this work \\
$P_{\rm puls}$ & $55.4734\pm0.0074$ & d & this work \\
$f_{\rm puls}/f_{\rm LSP}$ & $7.9970\pm0.0024$ & --- & this work \\
$K_1$ & $2.32^{+0.24}_{-0.23}$ & km\,s$^{-1}$ & this work$^\ast$ \\
$e$ & $0.31^{+0.11}_{-0.12}$ & --- & this work$^\ast$ \\
$\omega$ & $233^{+27}_{-21}$ & $^\circ$ & this work$^\ast$ \\
$\gamma$ & $-33.38\pm0.17$ & km\,s$^{-1}$ & this work$^\ast$ \\
$T_0$ & HJD\,$2\,457\,314^{+30}_{-20}$ & --- & this work$^\ast$ \\
Binary mass function & $4.8^{+1.8}_{-1.4}\times10^{-4}$ & ${\rm M}_\odot$ & this work$^\ast$ \\
\enddata
\tablecomments{$^\ast$ Assuming that {\it Gaia} FPR RV data for CZ\,Tuc can be modeled with a Keplerian orbit.}
\end{deluxetable}

\section{Discussion} \label{sec:discussion}

\subsection{Chance of random alignment}\label{sec:random_alignment}
The fact that the dominant pulsation frequency in CZ\,Tuc almost perfectly coincides with a harmonic of the LSP frequency could, in principle, be attributed to a very rare but entirely random alignment between the two signals. In such a scenario, the pair of frequencies would have no direct connection to each other. Since we know $f_{\rm puls}$ along with its uncertainty, we can estimate the following conditional probability. Let us assume that a particular self-excited mode has a frequency somewhere between $7.5f_{\rm LSP}$ and $8.5f_{\rm LSP}$. If all frequencies within this interval are equally probable, then the probability that the frequency falls within the $\pm1\sigma$ range around $8f_{\rm LSP}$ is $\sim16\sigma_{f_{\rm LSP}}/f_{\rm LSP}\approx 4\times10^{-3}$. This is therefore an unlikely situation for CZ\,Tuc, especially given the limited number of LSPs with $V\lesssim9$\,mag. On the other hand, if LSP variables (or at least some of them) are close binaries, the TEO offers a natural explanation of this intriguing ``coincidence''. Under such circumstances, the LSP of CZ\,Tuc automatically becomes its orbital period, with an unseen object inducing the observed TEO.

\subsection{Candidate modes for the TEO and their predicted amplitudes}\label{sec:candidate_modes}

\begin{figure}[ht!]
\includegraphics[width=0.47\textwidth]{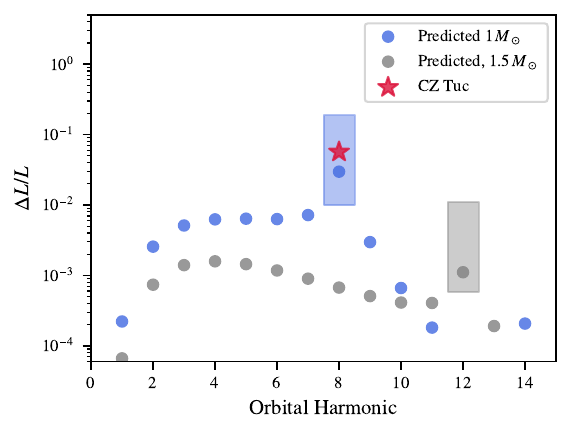}
\caption{Predicted tidally excited luminosity fluctuations vs. orbital harmonic, for a $1\,{\rm M}_\odot$ stellar model (blue points) and a $1.5\,{\rm M}_\odot$ stellar model (gray points) with similar luminosities. The shaded boxes indicate the range where the largest amplitude oscillation will lie 90\% of the time for each model. The red star is CZ\,Tuc's observed pulsation frequency and amplitude.}
\label{fig:TEO_Model}
\end{figure}

To determine whether the coherent pulsation signal that we observe in CZ\,Tuc could be an $N=8$ TEO, we predicted TEO amplitudes following the methodology of \cite{fuller2017}. We generated stellar models using MESA (\citealt{paxton2011,jermyn2023}, version 24.08.1) choosing masses of $1$ and $1.5\,{\rm M}_\odot$. We evolved these up the red giant and asymptotic giant branches until their luminosities were $L \approx 2\,500 \, {\rm L}_\odot$ and radii were $R \approx 150 \, {\rm R}_\odot$, comparable to estimated parameters of CZ\,Tuc (see Appendix~\ref{app:sed}). We then computed their spherical-harmonic degree $l=2$ oscillation modes using GYRE \citep{townsend2013}. These provided oscillation mode frequencies $\omega_\alpha$, damping rates $\gamma_\alpha$, tidal overlap coefficients $Q_\alpha$, surface luminosity fluctuations $L_\alpha$, and visibility coefficients $V_{lm}$ as defined in \cite{fuller2017}. The inferred orbital separation and eccentricity (see Appendix~\ref{app:rv}) were used to compute the tidal amplitude $\epsilon$ and tidal forcing coefficients $X_N$ at each integer multiple $N$ of the orbital frequency. We considered a non-rotating star, so the forcing frequencies are $\omega_N = N \Omega_{\rm orb}$. We assumed a companion mass $M_2 = 0.1 \, M_\odot$ and viewing inclination $i= 70^\circ$ relative to the orbital axis, consistent with the measured RV amplitude.

The amplitude of a TEO's luminosity fluctuation is
\begin{equation}
    \frac{\Delta L}{L} = \epsilon X_N V_{l m} |Q_\alpha L_\alpha| \frac{\omega_{N}}{\sqrt{(\omega_N - \omega_\alpha)^2 + \gamma_\alpha^2}} \, .
\end{equation}
In our case, the largest TEO will typically occur at frequencies $\omega_N$ nearly resonant with the star's $l=m=2$ f-mode, where $m$ denotes the azimuthal order. This is strongly dependent on the resonant detuning $\omega_N - \omega_\alpha$, which cannot be accurately predicted. The frequency difference between the f-mode and the nearest orbital harmonic can be as large as $\Omega_{\rm orb}/2$, or as small as zero, but 90\% of the time we expect $|\omega_N - \omega_\alpha|$ will lie in the range $0.025 \, \Omega_{\rm orb}$ - $0.0475 \, \Omega_{\rm orb}$. We thus used this range of detuning to calculate a 90\% confidence interval for the amplitude of the TEO at the value of $N$ closest to the f-mode frequency.

Figure~\ref{fig:TEO_Model} shows the results of our calculations for each model. The most important result is that it is possible for the observed pulsation to be a TEO at the eighth orbital harmonic for the $1\,{\rm M}_\odot$ model. For that model, the $l=2$ f-mode has a frequency close to the tidal forcing at $N=8$, causing the largest TEO to occur at that frequency. For the $1.5\,{\rm M}_\odot$ model, the larger envelope mass leads to a higher f-mode frequency, corresponding to $N=12$ and incompatible with the observed TEO candidate. The same would be true for any higher mass stellar models, smaller radius models, or models rotating in the prograde direction.

For our $1\,{\rm M}_\odot$ model, the observed pulsation amplitude is within the 90\% confidence interval, meaning it can be explained as a TEO due to the star's $l=2$ f-mode.
However, it is worth noting that the $l=2$ f-modes of our models have frequencies close to the $l=0$ f-modes (within $\sim 10\%$ in frequency). This means that stellar models that could account for the candidate TEO via a resonantly excited $l=2$ f-mode would also have an $l=0$ f-mode at a similar frequency. Since $l=0$ f-mode and low-overtone acoustic modes are the main sources of variability in AGB stars, we cannot discount the possibility that this is the source of a self-excited pulsation at $P_{\rm puls}\approx 55.5$\,d in CZ\,Tuc.

\subsection{Is CZ\,Tuc an isolated case?}\label{sec:other_teos}

\begin{figure}[ht!]
\includegraphics[scale=1]{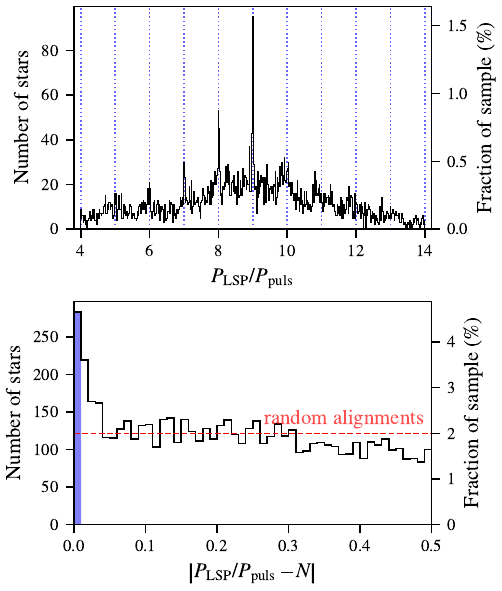}
\caption{Distribution of the period ratios $P_{\rm LSP}/P_{\rm puls}$ for the 6\,075 OGLE-III LSP stars described in Sect.~\ref{sec:other_teos}. (top) Histogram of the period ratios in bins of 0.02 centered on integer values, which are marked with vertical dotted lines. (bottom) Histogram of the distance between the period ratio and the nearest integer $N$, in bins of 0.01. The blue bar marks the stars with $|P_{\rm LSP}/P_{\rm puls}-N|<0.01$, and the red dashed line indicates the number of stars per bin expected for random alignments.}
\label{fig:pratios}
\end{figure}

If the dominant pulsation mode in CZ\,Tuc is indeed a TEO, and LSP stars are predominantly binary systems, it is reasonable to ask whether TEOs also occur in other variables of this type. This question is even more justified because our theoretical estimates for CZ\,Tuc (Sect.~\ref{sec:candidate_modes}) suggest that TEOs may reach relatively high amplitudes. Perhaps CZ\,Tuc is simply one of many LSPs in which at least one of the pulsation modes is tidally excited. To shed some light on this issue, we carried out the following preliminary experiment.

Our goal was to identify candidate TEOs in LSP variables by searching the OGLE-III catalog of LPVs in the Magellanic Clouds \citep{Soszynski2009,Soszynski2011} for near-integer period ratios (i.e., potential values of $N$ other than unity). The catalog contains 8\,904 stars classified as LSP variables, for each of which three periods are reported, $P_1$, $P_2$, and $P_3$. The first of these ($P_1$), corresponding to the strongest signal, represents LSP. We therefore examined whether the ratio of $P_1$ to either $P_2$ or $P_3$ falls within the range 4\,--\,14, motivated by the fact that the median ratio of $P_{\rm LSP}/P_{\rm puls}\approx 9$ for LSPs in general. This initial criterion is satisfied by 6\,075 stars. We then asked how often $P_1/P_{2,3}$ lies closer to the integer value than 0.01. We find that this is the case for 283 stars, corresponding to $4.7\%$ of the sample, while only up to 2\% can be random alignments (Fig.~\ref{fig:pratios}). The excess is the most pronounced for $N=7$, 8, and 9. On the other hand, Fig.~\ref{fig:pratios} shows no pronounced excess at $N\geq10$. This is in line with our theoretical predictions for CZ\,Tuc (Fig.~\ref{fig:TEO_Model}) -- for the $1\,{\rm M}_\odot$ model, the expected TEO amplitude drops to $\Delta L/L\lesssim10^{-3}$ at $N\geq10$, about an order of magnitude below the values predicted for $2\leq N\leq9$. Although these calculations were performed for the parameters of CZ\,Tuc, a similar decline of TEO amplitudes at high orbital harmonics is expected for other LSP stars with comparable masses and orbits, which would make such TEOs much harder to detect. The discussed excess of targets with $P_{\rm LSP}/P_{\rm puls}$ ratios close to an integer suggests that $\sim2.5$\,\% of them indeed include LSPs with TEOs. Hence, they constitute strong evidence in favor of the binary scenario. Because the chance of exciting a TEO increases with eccentricity, these would also be the LSPs with statistically the most eccentric orbits. It therefore appears that the case of CZ\,Tuc is not an isolated ``lucky'' example.

\section{Summary and conclusions}\label{sec:conclusions}
By compiling the data from the ASAS, Hipparcos, and DASCH surveys, we constructed the combined light curve for the bright LSP star -- CZ\,Tuc -- spanning as much as 80 years (Fig.~\ref{fig:combined_lc}). Two periodic variations can be identified for this target: an LSP with a period of $P_{\rm LSP}=443.62\pm0.12\,$d (Fig.~\ref{fig:phased_lcs}) and a distinct additional sinusoidal variation with a period of $P_{\rm puls}=55.4734\pm0.0074\,$d (Fig.~\ref{fig:isolated_teo}), which we interpret as a stellar pulsation period. Key parameters of CZ\,Tuc are presented in Table~\ref{tab:summary}. Most intriguingly, the ratio $P_{\rm LSP}/P_{\rm puls}=7.9970\pm0.0024$ is strikingly close to the integer value of 8.

This remarkable relationship can be interpreted in two ways. It could be a rare case in which, by pure coincidence, the pulsation frequency matches a harmonic of the LSP frequency, with no physical connection between the two. However, we consider this scenario unlikely given its probability ($\approx4\times10^{-3}$; Sect.~\ref{sec:random_alignment}), the precision of the frequency alignment, and limited number of bright LSP stars. In light of the prevailing hypothesis that the LSP phenomenon is rooted in binarity, a natural explanation is that the pulsation observed in CZ\,Tuc may be tidally excited by an unseen low-mass companion on an eccentric orbit.

Using theoretical calculations of tidally excited oscillations (TEOs), we demonstrated that the estimated physical and orbital parameters of CZ\,Tuc allow for production of its candidate high-amplitude TEO (Fig.~\ref{fig:TEO_Model}). In this interpretation, the observed stellar pulsation would correspond to a tidally excited quadrupolar fundamental oscillation mode of CZ\,Tuc.

Furthermore, although the excitation of a significant TEO requires specific conditions, CZ\,Tuc is not an isolated case. Among 6\,075 LSP stars from the OGLE-III catalog of LPVs in the Magellanic Clouds, 283 (4.7\%) have period ratios $P_{\rm LSP}/P_{\rm puls}$ within 0.01 of an integer, whereas only up to 2\% ($\sim120$ stars) are expected from random alignments (Fig.~\ref{fig:pratios}). This excess of about 160 stars is the most pronounced for $N=7$, 8 and 9, i.e., around the orbital harmonic of the candidate TEO in CZ\,Tuc, and it becomes much weaker at $N\geq10$, where our model for CZ\,Tuc predicts a sharp drop of TEO amplitudes. These results show that a few percent of LSP stars most likely host TEOs and are therefore candidate eccentric binary systems.

If the exceptional frequency tuning observed in CZ\,Tuc is indeed a signature of a TEO, this would provide another, yet completely independent argument in favor of the binarity of LSP stars (or at least some fraction of them). Since TEOs, by definition, can occur only in binary systems, their possible detection in the LSP light curves could constitute a pro-binarity argument just as strong as the detection of ``secondary eclipses'' in their mid-infrared light curves \citep{soszynski2021}. The 283 OGLE-III LSP stars with near-integer period ratios identified in this work are natural targets for follow-up studies aimed at verifying the TEO interpretation. LSP studies would also benefit from carrying out theoretical simulations of TEO amplitudes across the full range of possible configurations, in order to determine how common this phenomenon should be. Finally, the detection of TEOs in LSP stars would not only prove their binary status, but also provide evidence for non-zero eccentricities of their orbits \citep{decin2025}, which remain elusive to this day.

\begin{acknowledgments}
We are grateful to an anonymous referee, whose insightful comments helped us significantly improve the manuscript.

We are indebted to the OGLE collaboration for the use of facilities of the Warsaw telescope at Las Campanas Observatory, Chile, for their permanent support and maintenance of the ASAS instrumentation and to the Observatories of the Carnegie Institution of Washington for providing the excellent site for the observations. We would like to thank Mr. Wayne Rosing who has kindly allocated space and provided support for ASAS instruments at The Faulkes Telescope North site on Haleakala, Maui, Hawaii in the years 2006-2017. PKS, PI and DMS acknowledge support from ERC grant LSP-MIST (Grant Agreement No. 101040160). Funded by the European Union. Views and opinions expressed are however those of the authors only and do not necessarily reflect those of the European Union or the European Research Council. Neither the European Union nor the granting authority can be held responsible for them. This work has been funded by the National Science Centre, Poland, through grant 2022/45/B/ST9/00243 to IS.

This work has made use of data provided by Digital Access to a Sky Century @ Harvard (DASCH), which has been partially supported by NSF grants AST-0407380, AST-0909073, and AST-1313370. Work on DASCH Data Release 7 received support from the Smithsonian American Women's History Initiative Pool.

This work has made use of data from the European Space Agency (ESA) mission
{\it Gaia} (\url{https://www.cosmos.esa.int/gaia}), processed by the {\it Gaia}
Data Processing and Analysis Consortium (DPAC,
\url{https://www.cosmos.esa.int/web/gaia/dpac/consortium}). Funding for the DPAC
has been provided by national institutions, in particular the institutions
participating in the {\it Gaia} Multilateral Agreement.

This publication makes use of VOSA, developed under the Spanish Virtual Observatory (\url{https://svo.cab.inta-csic.es}) project funded by MCIN/AEI/10.13039/501100011033 through grant PID2023-146210NB-I00. VOSA has been partially updated by using funding from the European Union's Horizon 2020 Research and Innovation Programme, under Grant Agreement n$^\circ$ 776403 (EXOPLANETS-A).

\end{acknowledgments}




%
\facilities{ASAS, HIPPARCOS, Gaia}

\software{NumPy \citep{harris2020array}, Matplotlib \citep{Hunter:2007}}


\appendix

\section{Radial-velocity curve of CZ\,Tuc}\label{app:rv}
\begin{figure*}[ht!]
\includegraphics[scale=1]{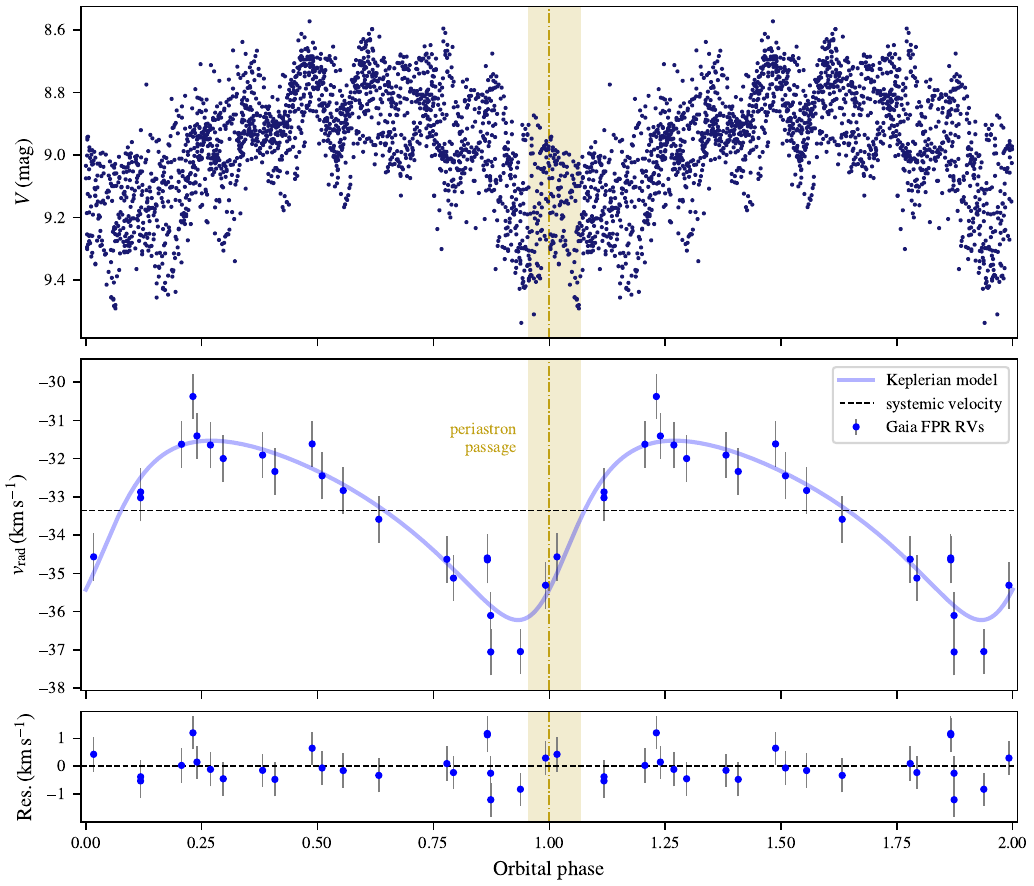}
\caption{Analysis of the RV data of CZ\,Tuc. The measurements are phase-folded using the LSP period of CZ\,Tuc, which is interpreted here as the orbital period. (top) The ASAS light curve phase-folded consistently with the RV curve. (middle) Fitting a Keplerian model to the RV curve from Gaia FPR. Individual RV measurements, along with their uncertainties, are shown as blue circles. The blue curve represents the best-fitting RV model. The horizontal dashed line indicates the systemic velocity. (bottom) The RV residuals from the best-fit. Phase zero corresponds to $T_0={\rm HJD}\,2\,457\,314$ and marks the time of periastron passage (vertical dash-dotted line). The vertical shaded region in all panels corresponds to the $\pm1\sigma$ uncertainty in $T_0$.}
\label{fig:rv-fit}
\end{figure*}

\begin{figure*}[ht!]
\includegraphics[scale=1]{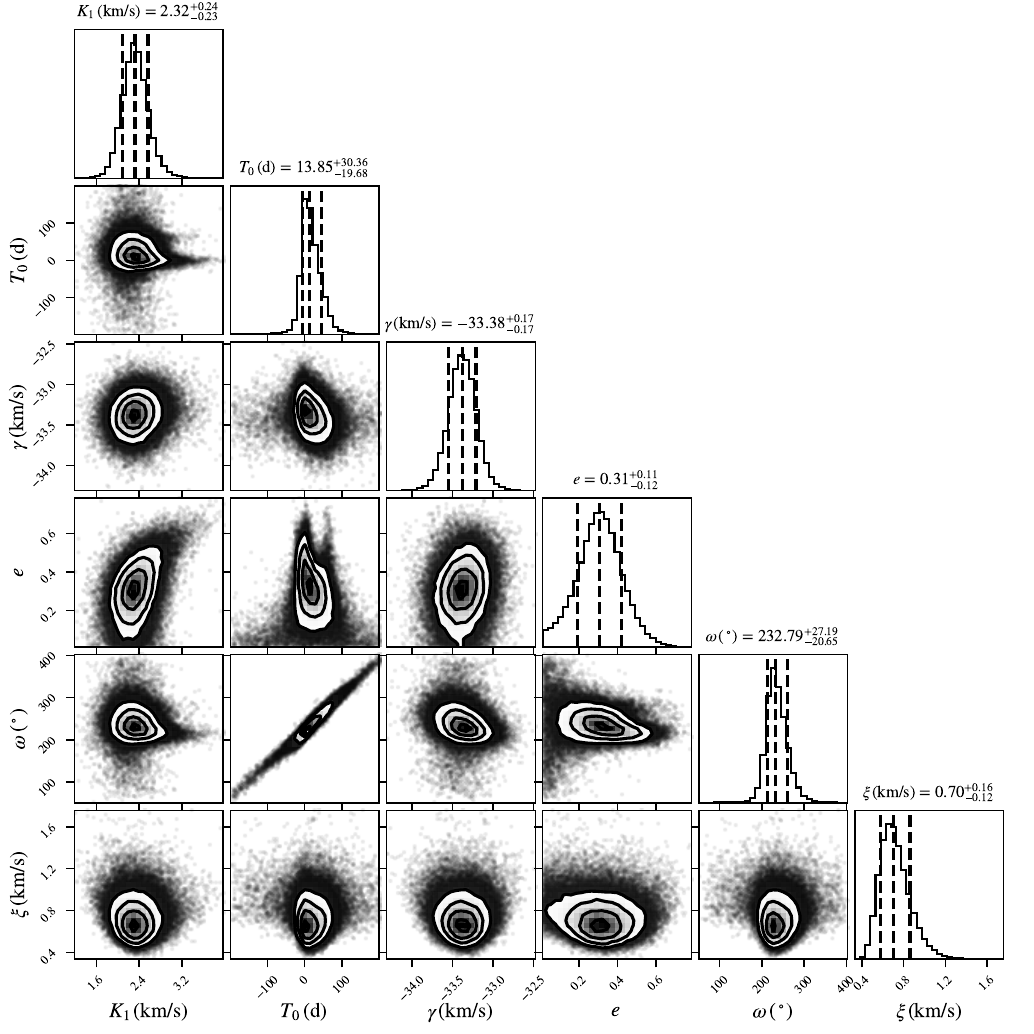}
\caption{Corner plot for the Bayesian fit of a Keplerian model to the RV data of CZ\,Tuc from Gaia FPR (cf. Fig.~\ref{fig:rv-fit}). The optimal parameter values, along with their uncertainties, are given above each of the one-dimensional histograms. The periastron passage time, $T_0$, corresponds to ${\rm HJD}-2\,457\,300$.}
\label{fig:rv-corner}
\end{figure*}

To determine the possible orbital parameters of CZ\,Tuc -- in particular the eccentricity of its orbit -- we fitted a Keplerian RV model to the {\it Gaia} FPR time-series data, which we retrieved directly from the VizieR service\footnote{\url{http://vizier.nao.ac.jp/viz-bin/VizieR-3?-source=I/361/vlpv}}. The RV curve contains 22 individual measurements with a typical uncertainty of $\sim0.2\,{\rm km\,s}^{-1}$. We performed a Bayesian analysis using the Python \texttt{emcee} package\footnote{\url{https://emcee.readthedocs.io/en/stable/}} \citep{2013PASP..125..306F}, with the following free parameters of the Keplerian model: RV semi-amplitude ($K_1$), orbital eccentricity ($e$), argument of periastron ($\omega$), time of periastron passage ($T_0$),  and systemic velocity ($\gamma$). We also marginalized the posterior over an additive, constant noise nuisance parameter ($\xi$), so that $\sigma_{{\rm RV},i}^2=\sigma_{{\rm Gaia},i}^2+\xi^2$. During fitting, the orbital period was fixed at the value of $P_{\rm LSP}=443.62$\,d obtained in Sect.~\ref{sec:LSP}. The resulting best-fitting RV model is shown in Fig.~\ref{fig:rv-fit}, while Fig.~\ref{fig:rv-corner} presents the corresponding corner plot together with the optimal values of the orbital parameters and their $\pm1\sigma$ errors. Most importantly, we find that the RV curve of CZ\,Tuc points toward a significantly eccentric orbit with $e=0.31^{+0.11}_{-0.12}$. The fitted parameters indicate a binary mass function of $4.8^{+1.8}_{-1.4}\times10^{-4}\,{\rm M}_\odot$, which, assuming a canonical primary's mass of $1\,{\rm M}_\odot$, translates to a minimal dynamical companion's mass of approximately $0.08\,{\rm M}_\odot$ and a semi-major axis of about $1.2$\,AU ($\sim260\,{\rm R}_\odot$).

\section{Estimate of the fundamental parameters of CZ\,Tuc}\label{app:sed}
Our theoretical predictions of the TEO amplitudes in CZ\,Tuc (Sect.~\ref{sec:candidate_modes}) required knowledge of several global parameters of this red giant. Unfortunately, we could not find any publicly available medium- or high-resolution spectra for this star. To avoid relying solely on broadband photometry, we therefore adopted the spectroscopic parameters reported by \cite{2025ApJ...984...58H}, who derived them from {\it Gaia} XP low-resolution spectra. These authors obtained $T_{\rm eff}\approx 3\,550$\,K, $\log(g/{\rm cm\,s}^{-2})\approx0.7$, and $[{\rm Fe/H}]\approx-0.3$. The quoted effective temperature and surface gravity are fully consistent with the spectral type of CZ\,Tuc, M2/3\,III.

Next, keeping the spectroscopic parameters fixed, we performed simplified modeling of CZ\,Tuc's SED using the online VOSA tool\footnote{\url{https://svo2.cab.inta-csic.es/theory/vosa/}} \citep[][version 7.5]{2008A&A...492..277B}. Using the \texttt{mwdust} package\footnote{\url{https://github.com/jobovy/mwdust}} \citep{2016ApJ...818..130B} and its built-in \texttt{Combined19} 3D dust maps \citep{2003A&A...409..205D,2006A&A...453..635M,2019ApJ...887...93G}, we estimated the Galactic contribution to the extinction toward CZ\,Tuc for $A_V\approx0.1$\,mag. To calculate forward SED models, we utilized the ATLAS9 grid of synthetic spectra \citep{2003IAUS..210P.A20C}. The best-fit solution we obtained corresponds to radius $R\approx140\,{\rm R}_\odot$ and luminosity $L\approx2\,700\,{\rm L}_\odot$. Given the purely approximate nature of our analysis, we did not perform formal error estimates.


\bibliography{references}
\bibliographystyle{aasjournalv7.1}



\end{document}